\documentclass[fontsize=9pt,letterpaper,twocolumn]{scrartcl}

\usepackage[lmargin=0.6in,rmargin=0.6in,tmargin=0.6in,bmargin=0.75in]{geometry}

\usepackage[utf8]{inputenc}
\usepackage[english]{babel}
\usepackage[T1]{fontenc}

\usepackage{amsmath,amssymb,amsfonts}
\usepackage[version=3]{mhchem} 
\usepackage{gensymb} 
\usepackage[low-sup]{subdepth}
\usepackage{MnSymbol}
\usepackage{csquotes}

\usepackage{times}
\usepackage[EULERGREEK]{sansmath}

\usepackage[colorlinks=true,
            linkcolor=linkcolor,
            urlcolor=linkcolor,
            citecolor=linkcolor,
            unicode,
            pdfencoding=auto]{hyperref}

\usepackage[
backend=biber,
safeinputenc,
style=nature
]{biblatex}
\usepackage{graphicx}
\usepackage{color}
\usepackage[dvipsnames,svgnames,x11names,hyperref]{xcolor}
\usepackage{float}

\usepackage[affil-it]{authblk}

\usepackage{caption}
\setcapindent{0em}
\makeatletter
\def\@maketitle{%
  \newpage
  \null
  \begin{center}%
  \let \footnote \thanks
    {\Huge \@title \par}%
    \vskip 1.5em%
    {\large
      \lineskip .2em%
      \begin{tabular}[t]{c}%
        \baselineskip=12pt
        \@author
      \end{tabular}\par}%
  \end{center}%
  \par
  \vskip 1.5em}
\makeatother
\setkomafont{section}{\normalfont\normalsize\sffamily\bfseries}
\setkomafont{subsection}{\normalfont\normalsize\sffamily}
\RedeclareSectionCommand[
  afterindent=false,
  beforeskip=1.5\baselineskip,
  afterskip=.2\baselineskip]{section}

\RedeclareSectionCommand[
   afterindent=false,
   beforeskip=1.5\baselineskip,
   afterskip=.2\baselineskip]{subsection}

\definecolor{linkcolor}{RGB}{6,69,173} 
\definecolor{diffcolor}{RGB}{175,31,36} 

\usepackage[
  colorlinks=true,
  linkcolor=linkcolor,
  urlcolor=linkcolor,
  citecolor=linkcolor,
  unicode,
  pdfencoding=auto
]{hyperref}

\usepackage[
]{physpack}

\newcommand{\kelvin}[0]{\ensuremath{\,\mathrm{K}}}
\newcommand{\tisetwo}[0]{{1\textit{T}-\ce{TiSe_{$2$}}}}
\newcommand{\pttisetwo}[0]{{1\textit{T}-\ce{Pt_{$x$}Ti_{$1-x$}Se_{$2$}}}}

\newcommand{\mysection}[1]{\section*{\uppercase{#1}}}
\newcommand{\mysubsection}[1]{\subsection*{#1}}

\title{%
Metal-to-insulator transition in Pt-doped \texorpdfstring{\ce{TiSe2}}{TiSe2} driven by\\
emergent network of narrow transport channels%
}

\author[1,2,3]{Kyungmin Lee}

\author[4]{Jesse Choe}

\author[5]{Davide Iaia}

\author[6]{Juqiang Li}
\author[6]{Junjing Zhao}

\author[7]{Ming Shi}
\author[7]{Junzhang Ma}
\author[7]{Mengyu Yao}

\author[5]{Zhenyu Wang}

\author[4]{Chien-Lung Huang}

\author[8]{Masayuki Ochi}

\author[9,10]{Ryotaro Arita}

\author[6]{Utpal Chatterjee}

\author[4]{Emilia Morosan}

\author[5]{Vidya Madhavan}

\author[1$\dagger$]{Nandini Trivedi}

\affil[1]{Department of Physics, The Ohio State University, Columbus, Ohio 43210, USA}
\affil[2]{Department of Physics, Florida State University, Tallahassee, Florida 32306, USA}
\affil[3]{National High Magnetic Field Laboratory, Tallahassee, Florida 32310, USA}
\affil[4]{Department of Physics and Astronomy, Rice University, Houston, Texas 77005, USA}
\affil[5]{Department of Physics and Frederick Seitz Materials Research Laboratory, University of Illinois Urbana-Champaign, Urbana, Illinois 61801, USA}
\affil[6]{Department of Physics, University of Virginia, Charlottesville, Virginia 22904, USA}
\affil[7]{Swiss Light Source, Paul Scherrer Institute, CH-5232 Villigen, Switzerland}
\affil[8]{Department of Physics, Osaka University, Toyonaka, Osaka 560-0043, Japan}
\affil[9]{Department of Applied Physics, University of Tokyo, Hongo, Tokyo 113-8656, Japan}
\affil[10]{RIKEN Center for Emergent Matter Science (CEMS), Wako, Saitama 351-0198, Japan}
\affil[$\dagger$]{email: trivedi.15@osu.edu}

\date{}

\begin{document}
\sansmath
\twocolumn[{\maketitle

\begin{abstract}
Metal-to-insulator transitions (MIT) can be driven by a number of different mechanisms, each resulting in a different type of insulator---%
Change in chemical potential can induce a transition from a metal to a band insulator;
strong correlations can drive a metal into a Mott insulator with an energy gap;
an Anderson transition, on the other hand, due to disorder leads to a localized insulator without a gap in the spectrum.
Here we report the discovery of an alternative route for MIT driven by the creation of a network of narrow channels.
Transport data on Pt substituted for Ti in \tisetwo{} shows a dramatic increase of resistivity by five orders of magnitude for few \% of Pt substitution,
with a power-law dependence of the temperature-dependent resistivity $\rho(T)$.
Our scanning tunneling microscopy data show that Pt induces an irregular network of nanometer-thick domain walls (DWs) of charge density wave (CDW) order,
which pull charge carriers out of the bulk and into the DWs.
While the CDW domains are gapped,
the charges confined to the narrow DWs interact strongly, with pseudogap-like suppression in the local density of states,
even when they were weakly interacting in the bulk, and scatter at the DW network interconnects thereby generating the highly resistive state.
Angle-resolved photoemission spectroscopy spectra exhibit pseudogap behavior corroborating the spatial coexistence of gapped domains and narrow domain walls with excess charge carriers.
\end{abstract}
\bigskip
}]

\mysection{Introduction}

\begin{figure*}[t]\centering%
\includegraphics{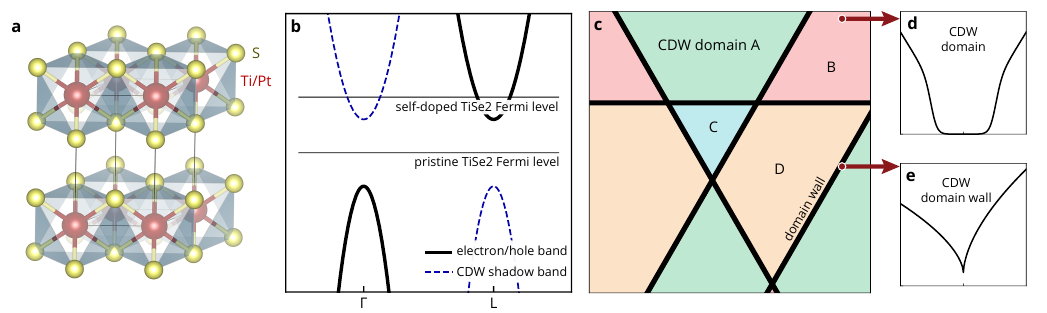}
\caption{\label{fig:1}
\textbf{Crystal structure and schematics.}
\textbf{a}~Crystal structure of \tisetwo{} and location of Pt when substituted.
\textbf{b}~Schematic band structure of \tisetwo{} with a valence band at the $\Gamma$ point and a conduction band at the L point.
The dashed lines indicate shadow bands due to CDW formation.
The as-grown sample is self-doped with the chemical potential within the conduction band.
The solid horizontal lines represent the chemical potential for the ``pristine'' system within the gap
and the ``self-doped'' system within the conduction band.
\textbf{c}~Pt substitution for Ti in \tisetwo{} creates domain walls between different charge density wave (CDW) ordered domains.
The CDW with ordering wavevector $\bfq = \mathrm{L}$ supports at least 4 types of domains in-plane, related to each other by lattice translations.
(Form factor of the CDW may introduce additional types of domains. See, e.g., Ref.~\cite{ishioka_prl_2010}.)
\textbf{d},~\textbf{e}~Schematic local density of states based on scanning tunneling spectroscopy results shown in Fig.~\ref{fig:STMFIG}g
inside a CDW domain (panel \textbf{d}), showing the semiconducting gap of the pristine \tisetwo{} band structure, and
on a CDW domain wall (panel \textbf{e}), showing a pseudogap with power-law behavior.
}
\end{figure*}

\begin{figure*}[t]\centering%
\includegraphics{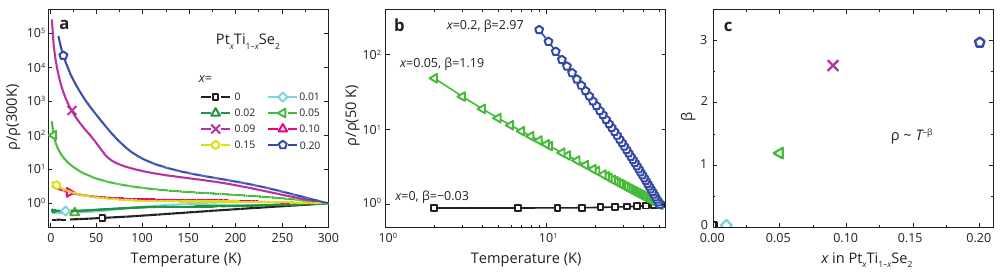}
\caption{\label{fig:transport}
\textbf{MIT in \pttisetwo{} from transport measurements.}
\textbf{a}~Resistivity ($\rho/\rho_{\text{300\,K}}$) as a function of temperature for various Pt doping $x$.
\pttisetwo{} initially shows metallic behavior but quickly becomes insulating with increasing $x$.
\textbf{b}~Log-log plot of resistivity vs. temperature for single crystal data at temperatures below 50\,K.
\textbf{c}~Power $\beta$  (in $\rho \sim T^{-\beta}$) vs. doping $x$, for single crystal measurements for the insulating samples.
}
\end{figure*}

\para{}
Pure \tisetwo{}, whose crystal structure is shown in Fig.~\ref{fig:1}a, undergoes a charge density wave (CDW) transition at $T_{\text{CDW}} =$ 200\,K with a $2\mathord{\times}2\mathord{\times}2$ charge order \cite{disalvo_prb_1976}.
While the nature of this CDW has been a matter of debate for decades, many of the recent experimental studies favor the electronically driven CDW
\cite{traum_prb_1978,cercellier_prl_2007,sugawara_acsn_2016,hildebrand_prb_2016,monney_prb_2016,kogar_s_2017,hedayat_prr_2019,mottas_prb_2019},
consistent with the theoretical description of excitonic insulator \cite{monney_prl_2011,chen_prl_2018}.
Nevertheless, a transport study has found finite resistivity in polycrystalline \ce{TiSe2} samples at low temperature \cite{chen_prb_2015}, consistent with metallic behavior.
The sensitivity of the low-temperature transport property on the synthesis condition \cite{moya_prm_2019} suggests that the observed metallic behavior in the polycrystalline sample is likely due to self-doping (see Fig.~\ref{fig:1}b).

\para{}
The temperature-dependent resistivity data on single crystals (Fig.~\ref{fig:transport}a) give a high-to-low temperature resistivity ratio $\rho(300K) / \rho(2K) = 10^{-\alpha}$ with an exponent $\alpha$ that is $\leq$ 0 for dopings $x \le 0.015$ (metallic behavior),
and $> 0$ (up to $\sim 5$) for $x \ge 0.02$ (insulating behavior),
indicative of MIT at around $x=0.015-0.02$ Pt doping with a remarkable $\sim 5$ orders of magnitude increase in the scaled resistivity compared to the $x=0$ compound.
[$x$ is an average nominal composition, which can vary locally within a crystal.]
Around this composition, the signatures of the CDW transition are also obscured by the diverging resistivity, consistent with the observation on the polycrystalline samples.
The temperature dependence of the resistivity in the insulating phase does not fit either the activated form or variable range hopping, $\rho \sim e^{(T_0/T)^{\beta}}$.
This is more clearly seen in the Zabrodskii plot \cite{zabrodskii_spjetp_1984} (Supplementary Note~1, Supplementary Fig.~1) where the zero slope suggests that the temperature dependence follows a power-law behavior,
without any exponential factor, as seen in Fig.~\ref{fig:transport}b and the doping dependence in Fig.~\ref{fig:transport}c.

\para{}
The high sensitivity of resistivity to Pt substitution as well as the unusual temperature-dependent resistivity point toward a highly unusual insulating state.
In this paper, we use a combination of electrical transport, angle-resolved photoemission spectroscopy (ARPES), and scanning tunneling microscopy (STM) to determine the mechanism for MIT in these compounds.
Guided by the STM and ARPES data,
we propose a model of narrow metallic channels on the domain walls of CDW, in an otherwise insulating background (See. Fig.~\ref{fig:1}),
which potentially provides explanations for the power-law temperature dependence of the resistivity, and points to an alternative path to obtain an MIT.

\mysection{Results and Discussion}

\mysubsection{Effects of Pt dopants}

\para{}
We find that Pt in \tisetwo{} has two effects:
(1) Our Hall measurements suggest that the Pt dopants introduce electron-like carriers to the system (see Supplementary Fig.~2).
Usually, such carrier doping of a semimetal/semiconductor drives the system towards a metallic phase;
this, however, is the opposite of what we find in \pttisetwo{}.
(2) Our DFT calculations find that it is energetically favorable for a Pt to substitute Ti rather than to intercalate.
(See Supplementary Note~3.)
Its $e_g$ orbitals have higher energy than the $t_{2g}$ orbitals of Ti, thus acting as a local potential impurity.
This Pt substitution is further corroborated by the shapes of the defects observed in topography images from scanning tunneling microscopy (shown in Supplementary Fig.~7)
and reaffirms the experimental finding on polycrystalline samples \cite{chen_prb_2015}.

\mysubsection{Pseudogap-like behavior observed in angle-resolved photoemission spectroscopy}

\begin{figure*}[t]\centering%
\includegraphics{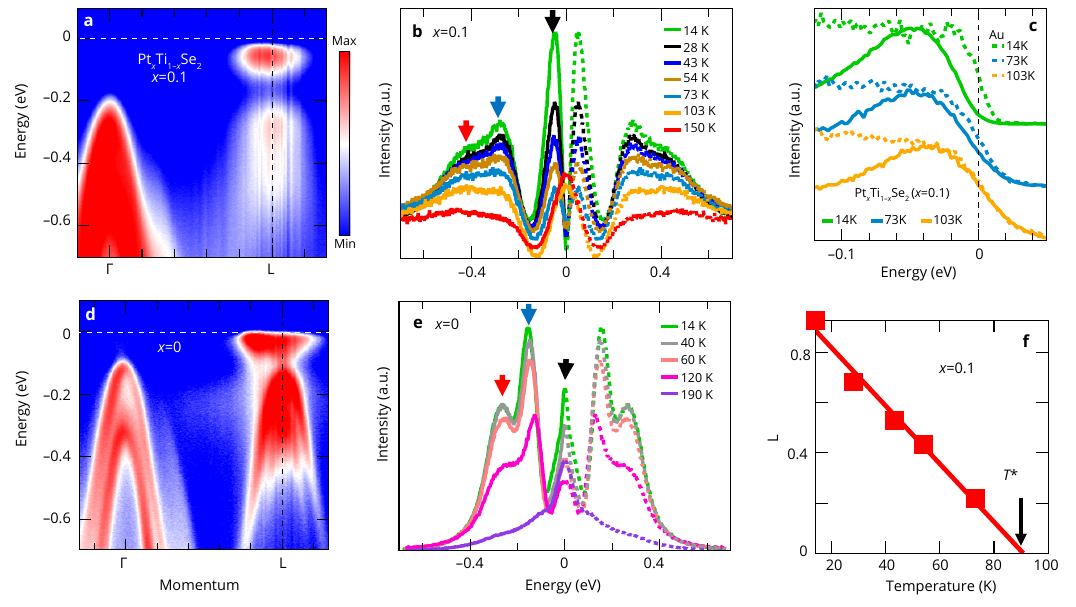}%
\caption{\label{fig:ARPES}
\textbf{Pseudogap-like suppression in ARPES spectra.}
\textbf{a}~ARPES energy-momentum intensity map (EMIM) for the $x=0.1$ sample at $T\sim14$\,K along $\Gamma$L direction.
\textbf{b}~Symmetrized EDCs as a function of temperature at the L point marked by the black dashed lines in \textbf{a} and \textbf{d}.
The positive-energy parts of the spectra are shown by dotted lines.
\textbf{c}~Comparison of the zoomed-in views close to $\mu$ of the leading edges of the EDCs from the $x=0.1$ sample and a polycrystalline gold sample at 14\,K, 73\,K, and 103\,K. The leading edges of the EDCs of the Pt-doped sample are shifted toward negative energy with respect to that of gold at 14\,K and 73\,K. This shift, and hence the energy gap, is absent at 103\,K.
\textbf{d} and \textbf{e}~are respectively the same as \textbf{a} and \textbf{b} but for $x=0$.
\textbf{f}~Temperature evolution of the filling up parameter $L(T) \equiv 1-\frac{I(T,0)}{I(T,\omega_{\text{peak}})}$ for the $x=0.1$ sample.
The black arrow points to the anticipated value of $T^*$, obtained from the extrapolation of $L$ to zero.
The red and blue arrows in \textbf{b} and \textbf{e} correspond to the CDW replica bands due to zone folding. Black arrows correspond to the spectral features associated with the conduction bands at the $L$ point.
}%
\end{figure*}

\begin{figure*}[t]\centering%
\includegraphics[width=360pt]{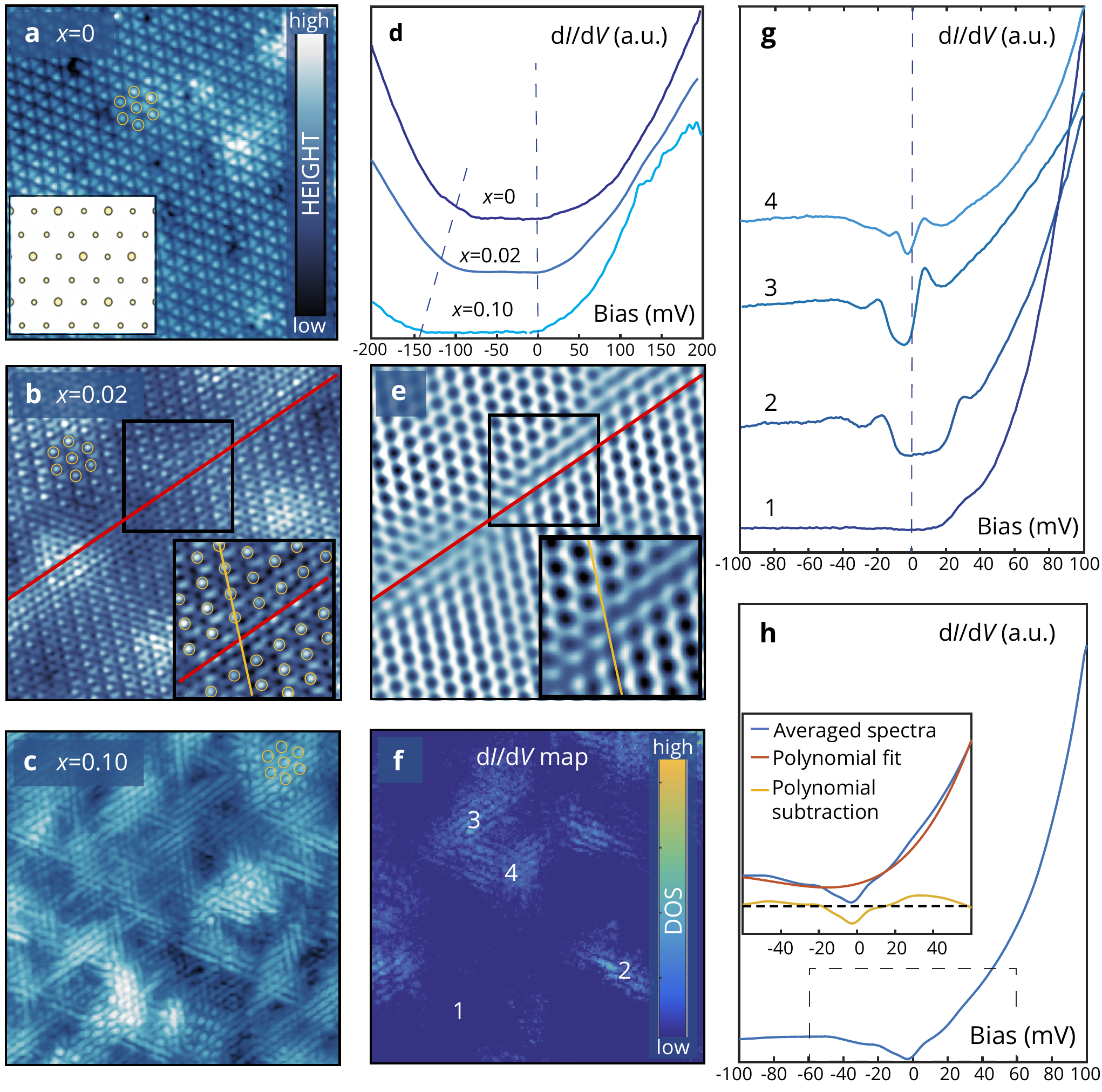}%
\caption{\label{fig:STMFIG}
\textbf{STM topography and tunneling spectra on Pt$_x$Ti$_{1-x}$Se$_2$.}
\textbf{a}~%
14\,nm $\times$ 14\,nm \ce{TiSe2} topography image acquired at 4\kelvin{}, $I = 20\,\mathrm{pA}$, and $V = 100\,\mathrm{mV}$.
The orange circles enclose CDW maxima.
The inset represents a sketch of the CDW order distribution in the top Se layer.
The large (small) yellow dots indicate CDW maxima (minima).
\textbf{b}~%
14\,nm $\times$ 14\,nm topography image of $x=0.02$ Pt-doped \ce{TiSe2} acquired at 4\,K, $I = 200\,\mathrm{pA}$, and $V = 50\, \mathrm{mV}$.
The red line indicates a CDW domain wall.
The orange circles enclose CDW maxima.
The inset is a magnified view of the region within the black square.
The orange line shows the $\pi$-shift between domains.
On one side of the domain wall, the orange line intersects the atoms while on the other side, it passes between the atoms.
\textbf{c}~%
14\,nm $\times$ 14\,nm topography image of $x=0.1$ Pt doped \ce{TiSe2} acquired at 4\,K, $I = 150\,\mathrm{pA}$, and $V = 100\,\mathrm{mV}$.
The orange circles enclose CDW maxima.
\textbf{d}~%
Most common spectra measured for $x=0, 0.02, 0.1$ doping.
\textbf{e}~%
Selective inverse Fourier transform of CDW peaks calculated from \textbf{b}.
The red line indicates the position of the CDW domain wall.
The inset is a magnified view of the area within the black box and the orange lines show the $\pi$-shift between domains.
\textbf{f}~%
$\rmd I/\rmd V$ image recorded at $V = -20\,\mathrm{mV}$ on the same area shown in \textbf{c}.
\textbf{g}~%
Averaged spectra measured on areas 1, 2, 3, and 4 of \textbf{f}.
\textbf{h}~%
$\rmd I/\rmd V$ spectra averaged over the entire area of \textbf{f}.
The inset shows the difference between the averaged spectrum and the polynomial fit.
In the topographies data, bright colors and dark colors correspond to higher and lower heights respectively,
while they correspond to higher and lower densities of states in $\rmd I/\rmd V$ data respectively.%
}%
\end{figure*}

\para{}
To gain insight into the MIT in \pttisetwo{}, we study the evolution of the single-particle spectral function of the doped single crystals from the metallic to the insulating regime using ARPES.
Figures~\ref{fig:ARPES}a and d show the energy-momentum intensity maps (EMIMs) \cite{UCR1-PDH} around the L point for $x=0.1$ and $x=0$, respectively.
  The EMIM depicts the ARPES intensity as a function of one of the in-plane momentum components and electronic energy $\omega$ referenced to the chemical potential $\mu$, while keeping the other orthogonal in-plane momentum component fixed.
Both samples show a hole-like band near the $\Gamma$ point and its CDW shadow near the L point.
Compared to that of the $x=0$ sample, the shadow band of the $x=0.1$ sample is much weaker and broader, indicating the loss of long-range CDW correlation with doping.
Interestingly, the $x=0.1$ sample shows suppression of the spectral weight at the Fermi level near the L point,
which we discuss next.

\para{}
The contrast between doped and undoped samples can be seen by examining the energy distribution curves (EDCs) in Figs.~\ref{fig:ARPES}b and e, which show ARPES data as functions of $\omega$ at a specific momentum.
To detect energy gaps near $\mu$, it is necessary to eliminate the effect of the Fermi function from the EDCs.
This can be approximately taken into account by employing symmetrization technique \cite{UCR2-JC-REVIEW, UCR3-ZX-REVIEW}.
A different method to accomplish this is to divide the EDCs by resolution-broadened Fermi function \cite{UCR2-JC-REVIEW, UCR3-ZX-REVIEW}.
In order to compare results obtained from these two different techniques,
we have conducted Fermi function division analysis of the EDCs from $x=0.1$ sample at various temperatures,
which are presented in Supplementary Fig.~6.
These are similar to the data in Fig.~\ref{fig:ARPES}b.
For $x=0$, the symmetrized EDCs (Fig.~\ref{fig:ARPES}e) exhibit a peak at $\mu$ at all measured temperatures implying the absence of any energy gap.
The data for the $x=0.1$ sample, (Fig.~\ref{fig:ARPES}b) is markedly different.
At higher temperatures of $T\ge$ 103\,K, the data show peaks at $\mu$, indicating the absence of an energy gap.
With decreasing temperature, however, the peaks of the symmetrized EDCs appear away from $\mu$, implying gapped electronic excitations.
It is to be noted that this energy gap is soft, meaning that there is finite spectral weight at $\mu$ even at the lowest measured temperatures.
This prompts us to refer to this energy gap as a pseudogap.
We note that this pseudogap can be independently verified from the STM data as well (Fig.~\ref{fig:STMFIG}h), as we will show later.

\para{}
The temperature evolution of the energy gap can also be investigated via leading-edge analysis of the ARPES data.
To this end, we focus on Fig.~\ref{fig:ARPES}c,
where we compare the leading edges of the $x=0.1$ sample and a polycrystalline gold sample at several temperatures (14\,K, 73\,K, and 103\,K).
Comparisons at several other temperatures (28\,K, 43\,K, 54\,K, and 150\,K) have been displayed in Supplementary Fig.~5 of Supplementary Note~4.
Collectively, these figures show that the leading edge of the EDCs of the Pt-doped sample ($x = 0.1$) is shifted towards negative energy with respect to that of gold at each measured temperature below 103\,K indicating the presence of an energy gap.
This shift disappears for $T\ge$ 103\,K, which evidences the absence of an energy gap.

\para{}
Interestingly, ARPES data suggest that the pseudogap does not ``close,'' rather it gets ``filled-up.''
This can be seen by tracking the peak positions of the symmetrized EDCs as a function of temperature.
For $14\kelvin{} \le T \le 73\kelvin{}$ we find that that the energy location of the peak of each symmetrized EDC is practically the same.
There is, however, a monotonic increase in the spectral weight at $\mu$ with increasing temperature from 14\kelvin{} onward.
Based on this, we speculate that it is the gradual increase in spectral weight at $\mu$ with increasing temperature that plays a dominant role in the eventual disappearance of the pseudogap at higher temperatures.
These observations are reminiscent of the temperature dependence of the pseudogap in underdoped cuprate high-temperature superconductors \cite{UCR4-MIKE-ARC, UCR5-NATCOM}.
We, however, do not suggest that the pseudogap has the same origin in these two very different systems.

\para{}
We can obtain an estimate for $T^\ast$ associated with the filling up of the pseudogap.
To do this, we define a parameter $L(T) \equiv 1-\frac{I(T,0)}{I(T,\omega_{\text{peak}})}$,
where $I(T,\omega)$ is the intensity of the symmetrized EDC at $\omega$, and $\omega_{\text{peak}}$ is the energy at the peak of the symmetrized EDC.
$I(T,0)$ is the intensity at $\mu$.
$T^\ast$ is then determined by the temperature at which $L$ vanishes \cite{UCR6-ZRTE3, UCR7-RAMAN}.
From Fig.~\ref{fig:ARPES}f, $T^\ast$ is estimated to be $\sim 90\kelvin{}$,
which is consistent with the observation of peaks at $\mu$ in the symmetrized EDCs for $T \ge 103\kelvin$ in Fig.~\ref{fig:ARPES}b as well as with the leading edge comparison with Au shown in Fig.~\ref{fig:ARPES}c.

\begin{figure}[t]\centering%
\includegraphics[width=252pt]{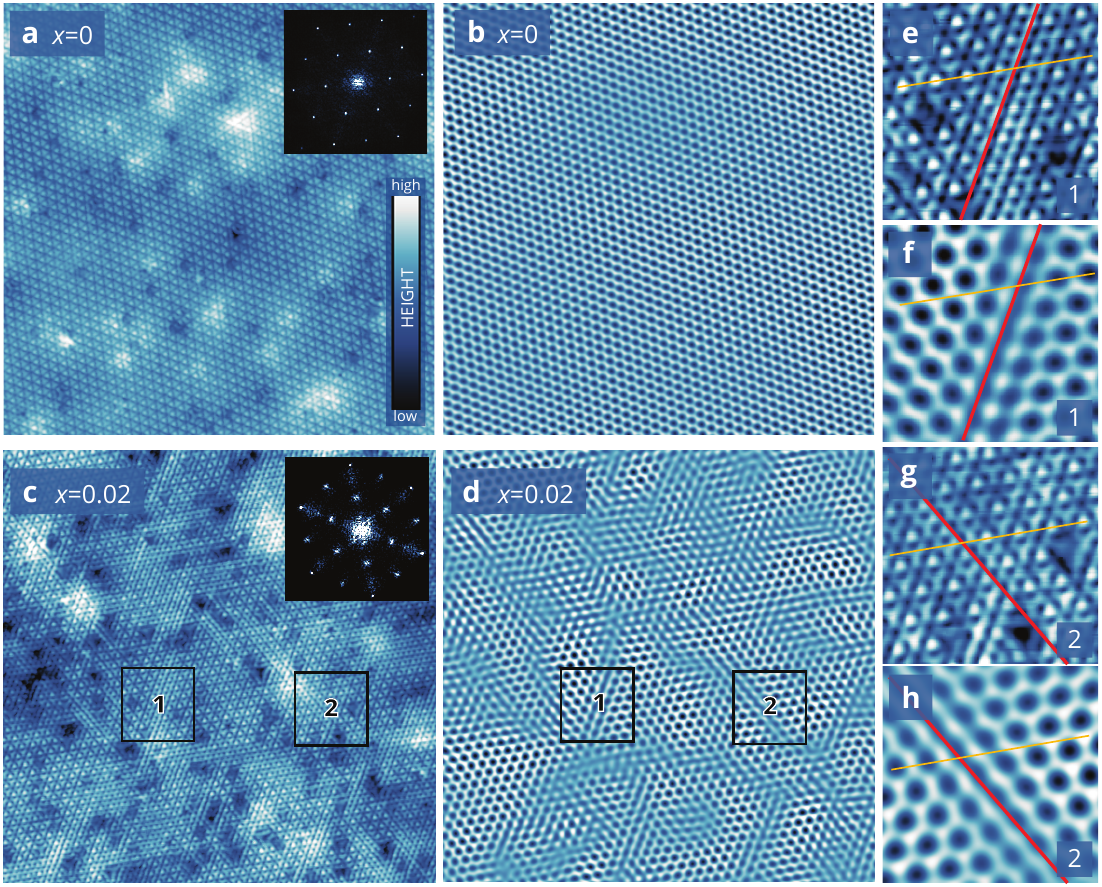}%
\caption{\label{fig:5}
\textbf{CDW domain walls.}
\textbf{a}~%
30\,nm\,$\times$\,30\,nm area recorded at 4\,K, 100\,mV and 20\,pA, for 0\,\% Pt doping.
The inset shows the Fourier transform of the topography.
\textbf{b}~%
Selective inverse Fourier transform of CDW peaks calculated from \textbf{a}. Consistent with previous STM data on pristine samples, the domain wall density is very low. In fact, there are no domain walls in this area.
\textbf{c}~%
30\,nm\,$\times$\,30\,nm area recorded at 300\,mK, 100\,mV and 20\,pA, for 0.02\,\% Pt doping.
The inset shows the Fourier transform.
\textbf{d}~%
Selective inverse Fourier transform of CDW peaks calculated from \textbf{c}.
A network of domain walls (seen as short streaks in the principle crystalline directions) is visible.
\textbf{e}-\textbf{h}~%
Magnified views of the smaller rectangular areas shown in \textbf{c} and \textbf{d} indicated as 1 and 2 are shown in (\textbf{e} and \textbf{f}) and 2 (\textbf{g} and \textbf{h}), respectively.
The red lines indicate CDW domain walls.
The orange lines indicate the $\pi$-shift between two CDW domains.
The phase shift can be seen clearly in \textbf{e} and \textbf{g} by tracking the positions of the atoms across the domain walls.
}
\end{figure}

\para{}
ARPES data, therefore, reveal crucial information on the opening of a pseudogap at the chemical potential,
which provides potential clues to the dramatic increase in resistivity with Pt-substitution,
but also leaves us with some unanswered questions on the origin of the gap.
Since the chemical potential shifts up into the conduction band with doping, a transition to a band insulator cannot be the origin of the gap.
CDW is also unlikely to be the cause since the ordering is stronger in the undoped sample which is gapless at the chemical potential.
In addition, we do not see evidence of hybridization of the conduction band with another band separated by the CDW wavevector.
On the other hand, the fact that the gap fills up with increasing temperature suggests that it is a correlation-driven gap,
which raises the question of why correlation effects in Pt-substituted \ce{TiSe2} are so different from the parent \ce{TiSe2}.

\mysubsection{Domain wall network revealed by scanning tunneling microscopy}

\para{}
To determine the origin of the pseudogap in \pttisetwo{}, we turn to local properties using STM.
\pttisetwo{} single crystals were cleaved at 80\kelvin{} before being inserted into the STM head held at 4\,K.
From STM topography shown in Figs.~\ref{fig:STMFIG}a-c (also see Supplementary Note~5, Supplementary Figs.~7--9), we find that the sample becomes increasingly disordered with the Pt doping,
although the CDW survives as a short-range ordered state (Supplementary Fig.~9).
Through selective inverse Fourier transform (explained in Supplementary Note~6),
we can identify that the origin of the disorder is due to the increase in the domain wall density (see Fig.~\ref{fig:5}).
This is consistent with the ARPES observation that the back-folded band becomes broader and weaker with Pt-doping.

\para{}
The overall trend of the STM spectra with Pt doping can be seen in the typical STM spectra obtained far from impurities or domain walls for $x=0$, $0.02$, and $0.1$ samples as shown in Fig.~\ref{fig:STMFIG}d.
The U shape of the spectra is consistent with the band gap of a semiconductor with a flat bottom representing the bulk indirect band gap between the p- and d- states.
Intrinsic doping leads to the Fermi energy $\mu$ being positioned close to the bottom of the conduction band for each of these $x$ values.
A closer look at the density of states (Supplementary Fig.~11) indicates that for these pristine samples,
the Fermi energy lies slightly inside the conduction band, resulting in metallic transport behavior.
STM data show that the gap between the conduction and valence bands increases with increasing $x$
(dotted lines in Fig.~\ref{fig:STMFIG}d) which is consistent with ARPES data in  Figs.~\ref{fig:ARPES}a and d.

\para{}
We now focus on the behavior of the local density of states (LDOS) in the $x=0.1$ samples.
An interesting picture emerges from our study of  $\rmd I /\rmd V$ maps
(measured at  $V_{\text{bias}} = -20\,\mathrm{mV}$ in the same area as the topography in Fig.~\ref{fig:STMFIG}c)
that reveals regions with different density-of-states behaviors.
In the larger area region within the domains denoted as 1 in Figs.~\ref{fig:STMFIG}f and g the density of states is close to zero around $\mu$ with the conduction band bottom seemingly shifted to higher energies.
On the other hand, in the regions denoted by 2, 3, and 4, spatially correlated with the location of the domain walls as identified by the topography,
we observe a finite density of states around $\mu$ with a small pseudogap.
This suggests that there is a charge or spectral weight transfer from the domains to the domain walls.

\para{}
The spatially averaged spectrum shown in Fig.~\ref{fig:STMFIG}h shows pseudogap,
i.e., a depletion in the density of states relative to background on an energy scale of 30\,meV around $\mu$,
which is highlighted when a polynomial background is subtracted
(see inset to Fig.~\ref{fig:STMFIG}h as well as Supplementary Fig.~10).
Of importance is our observation that the energy scale of the pseudogap is comparable to the one observed by ARPES.
Similar pseudogap features are not visible in the averaged spectra of the $x=0$ samples (Supplementary Fig.~10).
Based on this we can conclude that the pseudogap feature for $x=0.1$ sample can be independently observed from STM and the associated energy scales qualitatively agree with the ARPES observations.
In addition, STM provides a detailed local view that the Pt dopants disrupt the CDW and affects the local electronic spectra (see also Supplementary Note~5, Supplementary Figs.~7--9).
It is thus important to understand the role of the Pt dopants theoretically.

\para{}
It is encouraging that the pseudogap measured in ARPES agrees with the pseudogap in the averaged density of states;
it is, however, unclear how with a finite density of states at zero energy,
the resistivity in the Pt-substituted samples could be 5 orders of magnitude larger than the as-grown samples.

\mysubsection{Emergent network of narrow transport channels}

\para{}
The substituted Pt acting simply as scattering centers for the charge carriers does not explain the extreme sensitivity of transport property to doping.
The Pt dopants, however, can provide a pinning potential for the commensurate CDW.
Following the argument by Imry and Ma~\cite{imry-prl-1975}, two dimensions is the lower critical dimension for commensurate order with a random field.
This suggests that in \ce{TiSe2}, which is a layered system with weak interlayer coupling, Pt dopants can strongly disrupt the long-ranged CDW order.
This is consistent with the large heterogeneity in the LDOS measured by STS in the doped samples (Figs.~\ref{fig:STMFIG}c, f, and g), and also provide a possible explanation for the dramatic impact of doping on transport.

\para{}
The large resistance increase and the transition to an insulating phase suggest a model in which the introduction of Pt in \ce{TiSe2} produces an irregular network of channels as shown by the STM images.
Our DFT calculations indicate a large potential difference between Ti and a Pt impurity that substitutes for Ti.
The effect of such an impurity on a CDW material is akin to a stone impacting a glass surface and producing cracks.
While the understanding of the precise mechanism for the creation of domain walls is left for a future study,
our results paint a picture of the consequences of having domain walls after Pt substitution.
The STM LDOS indicates that the charge within the self-doped \ce{TiSe2} moves into the domain walls creating a large full gap inside domains and a weaker pseudogap within the domain wall.
Thus dc transport occurs primarily via the network of domain walls with the resistance being dominated at the interconnects.

\para{}
The power-law $\rho(T)$ has previously been observed in (quasi-) one-dimensional systems
\cite{bockrath_n_1999,yao_n_1999,bachtold_prl_2001,kanda_prl_2004,gao_prl_2004,monteverde_prl_2006,coiffic_apl_2007,dayen_epjap_2009, zaitsev-zotov_jpcm_2000,chang_rmp_2003,tserkovnyak_prb_2003,slot_prl_2004,aleshin_prl_2004,rahman_jpcm_2010,venkataraman_prl_2006,zhou_apl_2007}
and was interpreted as due to Luttinger liquid physics.
In such a scenario, the power-law suppression of the LDOS should be due to the tunneling of an electron into a system in which the elementary particles are not electrons but fractionalized charge and spin bosonic collective modes.
In spite of the suppressed LDOS, the resistance of a clean Luttinger channel vanishes at zero temperature \cite{kane_prb_1992}, except for the contact resistance.
This is a singular example of the role of vertex corrections in strongly modifying the transport behavior from that reflected in the single-particle Greens function and LDOS.
The observed resistivity $\rho \sim T^{-\beta}$ in this case then should arise from strong scattering at the interconnects of the network of Luttinger channels created in Pt-substituted \ce{TiSe2}---%
a Luttinger liquid junction formed between intersecting channels can lead to a diverging resistance \cite{aristov_prb_2017}.

\para{}
Networks of metallic domain walls have previously been observed in
Cu-intercalated TiSe$_2$ \cite{yan_prl_2017}, and also in other contexts:
integer quantum Hall systems \cite{chalker_jpcssp_1988},
magnetically ordered systems \cite{ma_science_2015},
nearly commensurate charge density wave systems \cite{park_nc_2019},
and also recently in moir\'e systems \cite{huang_prl_2018,efimkin_prb_2018,sunku_s_2018,xu_nc_2019} with periodic networks determined by the moir\'e pattern.
Remarkably, power laws have also been observed in the mass transport through a network of dislocations in solid $^4$He generated by a pressure difference \cite{shin_prb_2019},
suggesting a more ubiquitous occurrence of the network of narrow channel transport.
The mechanism for the generation of networks is system-specific---disorder, quench, dislocation, etc.
Nevertheless, the transport properties at temperatures below the domain gap scale are determined by the electronic structure of the network, as well as its connectivity.
A recent theoretical work shows that a network of Luttinger channels supports a stable insulating fixed point with power-law temperature dependence \cite{lee_2020_arxiv}.

\para{}
Let us point out that Luttinger liquid network is not the only explanation for the power-law-like temperature dependence.
Another possible mechanism for an insulating behavior with power-law temperature dependence is the ``rare chain hopping,''
\cite{pollak_prl_1973,tartakovskii_sps_1987,glazman_spj_1988,levin_sps_1988,rodin_prl_2010,rodin_prb_2011},
which describes the transport behavior of percolating channels whose resistivity is controlled by the connectivity to the network.
It leads to a much weaker $T$ dependence compared to variable range hopping, which may appear as a power law in a limited temperature range.
Nevertheless, the formation of one-dimensional conduction channels is a crucial ingredient in the rare chain hopping model.

\para{}
In summary, through a collective study of transport, ARPES, and STS,
we argue that the low energy insulating behavior in Pt-doped TiSe$_2$ arises as a result of an interplay between electron correlation and disorder,
and is governed by an emergent network of CDW domain walls behaving as narrow channels of low energy transport (see Figs.~\ref{fig:1}c-e).

\para{}
Going further, it would be interesting to study whether the MIT mechanism discussed here applies to other systems with CDW.
Immediate questions specific to TiSe$_2$ are how unique is the effect of Pt, and whether other substitutions act similarly.
Our preliminary investigations with other substitutions indicate a much smaller increase of resistivity, by less than an order of magnitude, as opposed to the five orders of magnitude increase with Pt.

\para{}
More broadly, our finding that the CDW domain walls provide one-dimensional channels of transport has implications for other CDW systems.
In the Cu-intercalated \ce{TiSe2} where superconductivity and the incommensurate CDW are observed around the same doping \cite{morosan_superconductivity_2006},
the domain walls could play a major role by channeling electrons into these narrow channels and thereby enhance correlation effects for superconductivity and magnetism.
As was pointed out in Ref.~\cite{chen-prb-2019},
when the superconductivity coexists with the CDW close to commensurate-incommensurate transition,
the network of domain walls can lead to percolative superconductivity.%

\mysection{Methods}

\mysubsection{First-principles calculation}

\begin{small}
\para{}
The first-principles band structure and the partial density of states (pDOS) of pure \ce{TiSe2} and PtSe$_2$ were calculated using the modified Becke-Johnson potential~\cite{mBJ1,mBJ2} as implemented in the {WIEN2k} package~\cite{wien2k}.
Spin-orbit coupling was not included in the calculation throughout this paper.
For simplicity, in these calculations, we used the experimental lattice constants $a=3.537$\,\AA \ and $c=6.007$\,\AA \ for TiSe$_2$~\cite{chen_prb_2015} and $a=3.7278$\,\AA \ and $c=5.0813$\,\AA \ for PtSe$_2$~\cite{PtSe2strct} with the fixed internal coordinate $c_{\mathrm{Se}}=0.25$ where Ti or Pt is placed on the $c=0$ plane.

\para{}
For the energy comparison between different Pt positions, we optimized the crystal structures using the DFT-D3 energy functional~\cite{DFTD3} as implemented in the Vienna ab initio simulation package~\cite{paw,vasp1,vasp2,vasp3,vasp4}.
Calculations for the cases (2) and (3) employed the $6\times 6\times 2$ supercell.

\para{}
For the calculation of the onsite energies of the Wannier orbitals,
we calculated the band structure of PtTi$_{17}$Se$_{36}$ using the $3\times 3\times 2$ supercell,
and then extracted the Wannier functions of the Pt-$e_g$ $+$ Ti-$t_{2g}$ model using the \textsc{WIEN2k}, \textsc{Wien2Wannier}, and \textsc{Wannier90} softwares~\cite{wannier1,wannier2,Wien2Wannier,Wannier90}.
We omitted the maximal localization procedure.
The crystal structure was fixed throughout the calculations for simplicity.

\end{small}

\mysubsection{Growth}

\begin{small}
\para{}
Single crystals of \ce{Pt_{$x$}Ti_{$1-x$}Se2} were grown by chemical vapor transport using excess Se as the transport agent.
Stoichiometric amounts of Pt and Ti, as well as Se power with approximately 50\% excess Se by mass were sealed in evacuated quartz tubes of approximately 6 inches in length and 0.5 inches in diameter.
Samples were heated in a gradient furnace for ten days minimum at a gradient of 900-1100$\,\degree$C then cooled to room temperature. Reported doping amounts in this study are for nominal compositions.
\end{small}

\mysubsection{ARPES}

\begin{small}
\para{}
Temperature-dependent ARPES measurements on undoped and Pt-doped \tisetwo{}  single crystals were performed using a Scienta R4000 electron analyzer at the SIS beamline of Swiss Light Source, Paul Scherrer Institute, Switzerland.
The measurements were performed using plane-polarized light with 45\,eV photon energies ($h\nu$). The energy and momentum resolutions were approximately 10--20\,meV and 0.0055\,$\mathring{\mathrm{A}}^{-1}$, respectively.
\end{small}

\mysubsection{STM}

\begin{small}
\para{}
The experiment was performed in an ultrahigh vacuum (UHV) system with a base pressure lower than 10$^{-10}$\,mbar and at a temperature of $\sim$4\,K.
The samples were cleaved at room temperature and immediately inserted into the STM scanner at 4\,K.
Differential conductance ($\rmd I / \rmd V$) spectra were acquired using a standard lock-in technique.
\end{small}

\mysection{Data Availability}

\begin{small}
All relevant data are available from the corresponding author upon reasonable request.
\end{small}

\mysection{Acknowledgements}

\begin{small}
K.L. and N.T. acknowledge support from National Science Foundation (NSF) Grant No.~DMREF-1629382.
J.C., C.L.H., and E.M. acknowledge support from NSF Grant No.~DMREF-1629374.
D.I., Z.W., and V.M. acknowledge support from NSF Grant No.~DMREF-1629068.
J.L., J.Z., and U.C acknowledge support from NSF Grant No.~DMREF-1629237.
M.S., J.M., and M.Y. were supported by the Sino-Swiss Science and Technology Cooperation Grant No.~IZLCZ2-170075.
R.A. acknowledges the financial support of Japan Society for the Promotion of Science Kakenhi Grant No.~16H06345.
\end{small}

\mysection{Competing Interests}
\begin{small}
The authors declare no competing interests.
\end{small}

\mysection{Author Contributions}
\begin{small}
K.L. and N.T. performed theoretical modeling.
J.C., C.L.H., and E.M. grew the crystals and performed the transport measurements.
D.I., Z.W., and V.M. performed the STM measurements.
J.L., J.Z., M.S., J.M., M.Y., and U.C. performed the ARPES measurements.
M.O. and R.A. performed the DFT calculations.
\end{small}

\printbibliography[title={REFERENCES}]


\end{document}


\maketitle

\section{Zabrodskii-Zinov'eva Analysis}

\begin{figure}[h]\centering
\includegraphics[width=240pt]{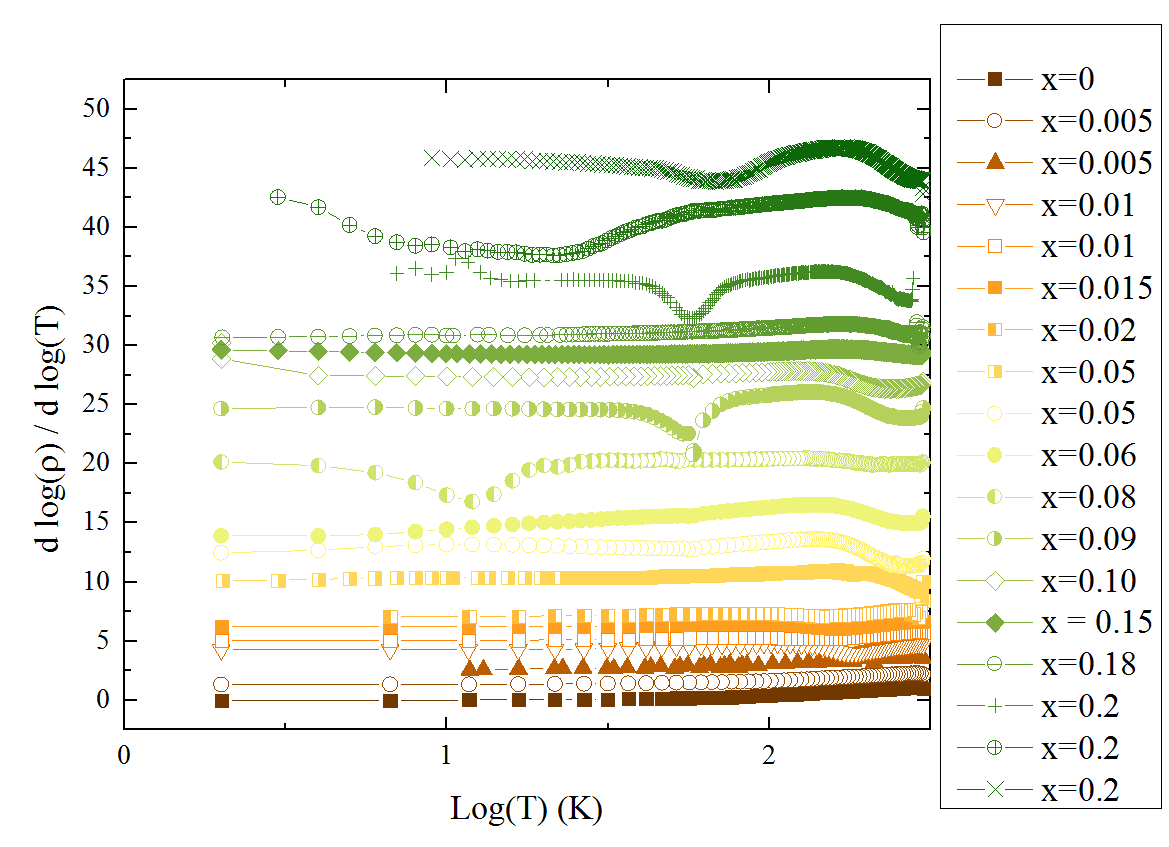}
\caption{\label{fig:zabrodskii}%
Zabrodskii-Zinov'eva plot for various dopings.
}
\end{figure}

\para{}
We apply the Zabrodskii-Zinov'eva analysis {\cite{zabrodskii_spjetp_1984}} to understand the temperature dependence of resistivity.
Assuming that the temperature dependence of the resistivity follows the following general form
\begin{align}
    \rho(T) = B T^{-m} \exp ( T_0 / T )^x,
\end{align}
investigating the temperature dependence of the following quantity
\begin{align}
w(T)
    &\equiv - \frac{\rmd \log \rho}{\rmd \log T}
    = m + x \left(\frac{T_0}{T}\right)^{x}
\end{align}
to extract the exponential factor.
As shown in Fig.~\ref{fig:zabrodskii}, $w(T)$ vs. $T$ in log-log scale appears nearly flat in all the samples we have measured, indicating the lack of exponential dependence.

\clearpage
\section{Hall Coefficients of Pure and Doped \texorpdfstring{$\mathrm{TiSe}_2$}{TiSe2}}

\begin{figure}[h]\centering
\includegraphics[width=260pt]{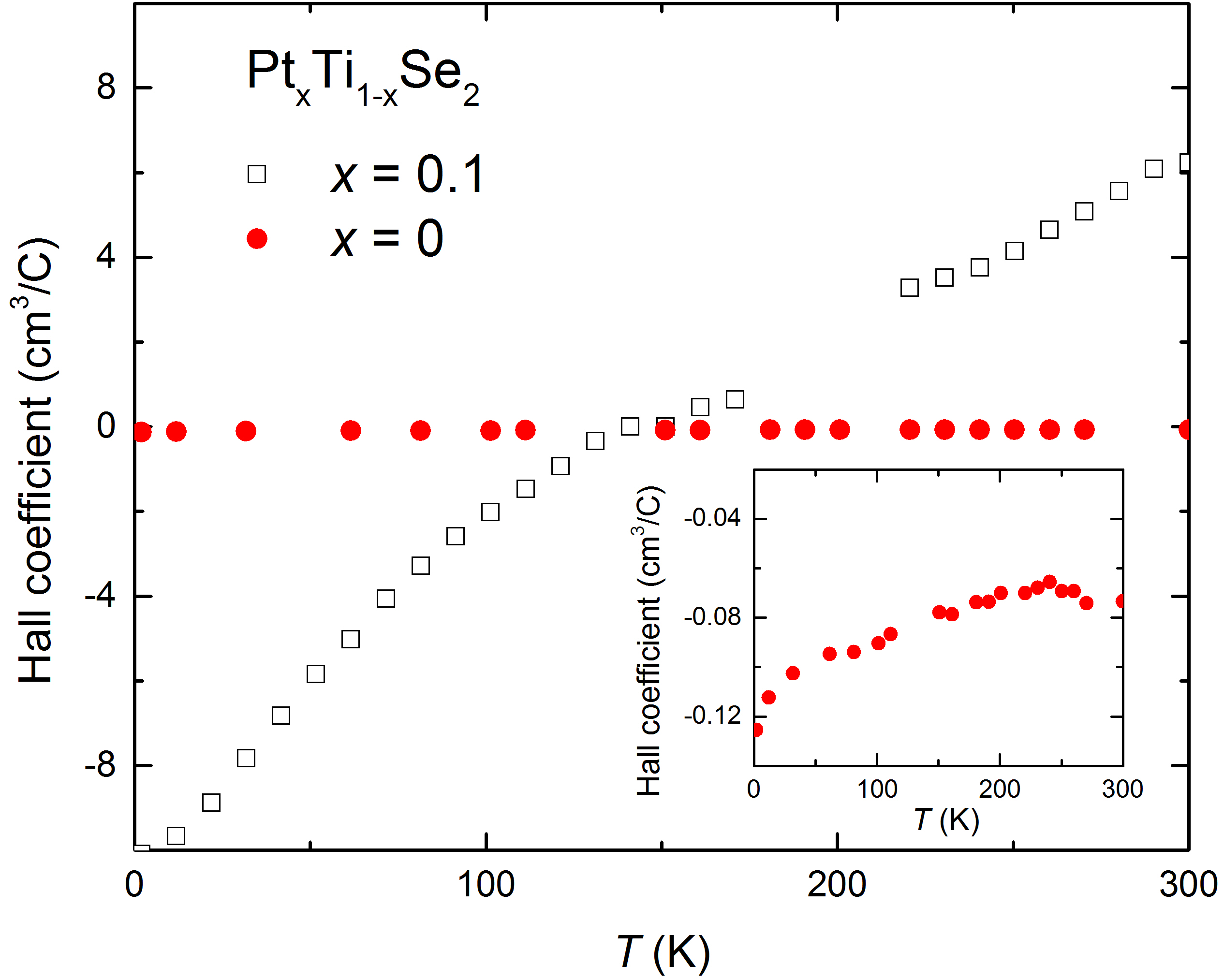}
\caption{\label{fig:Hall}Hall measurements for $x=0$ and $x=0.1$ samples.
\diff{}{The inset shows a close-up of the Hall coefficient for the undoped case.}
}
\end{figure}

We present Hall measurement data in Supplementary Fig.~\ref{fig:Hall} for an undoped and a doped sample.
Compared to the undoped ($x=0$) sample, which shows little Hall response,
the $x=0.1$ sample shows a negative Hall coefficient at low temperatures,
indicating that the Pt dopants introduce electron-like carriers.

\clearpage
\section{First Principles Calculation}

\begin{figure}[h]\centering%
\includegraphics{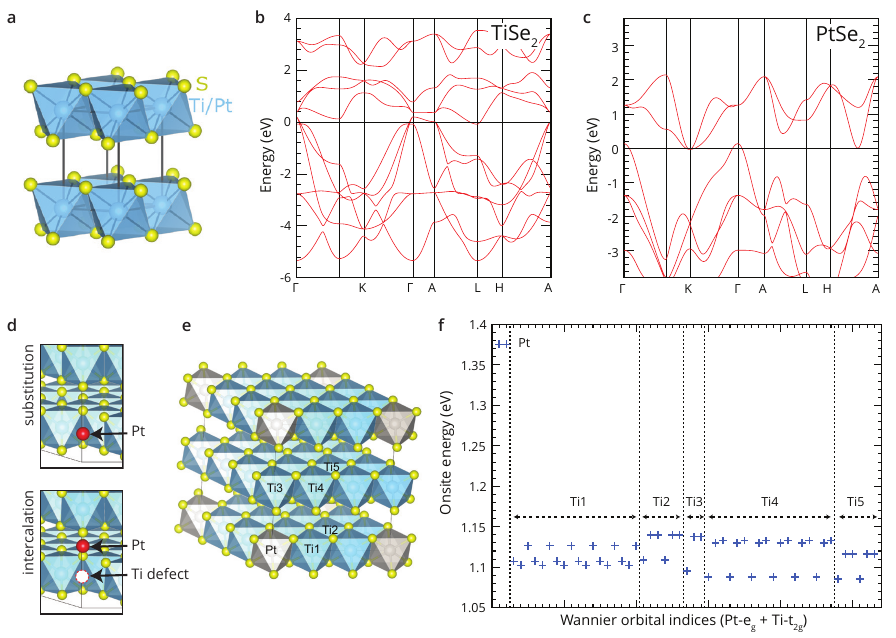}
\caption{\label{fig:2}%
\textbf{a}~Crystal structure of \tisetwo{} depicted using the VESTA software~\cite{VESTA}.
\textbf{b}~First-principles band structure of pure $1T$-\ce{TiSe2} in the absence of CDW superstructure.
\diff{\textbf{c}~Partial density of states (pDOS) of pure \ce{TiSe2}}{%
\textbf{c}~First-principles band structure of pure $1T$-\ce{PtSe2}.%
}
%
\textbf{d}~Crystal structure of \pttisetwo{} where Pt substitutes for Ti (\diff{left}{top}) and Pt is located at the intercalated position (\diff{right}{bottom}). 
\textbf{e}~Crystal structure of Pt-substituted \ce{TiSe2}.
\textbf{f}~Onsite energies of the Wannier functions in the Pt-$e_g$ $+$ Ti-$t_{2g}$ model.
}
\end{figure}

\para{Band structure}
\diff{%
Supplementary Fig.~\ref{fig:2}b presents the first-principles band structure and the partial density of states (pDOS) of pure \ce{TiSe2}.
Although the calculated band structure is semimetallic because we did not consider the CDW state using a larger supercell,
we can see that the conduction bands mainly consist of the Ti-$t_{2g}$ states.
The band structure of pure \ce{PtSe2} is also shown in Supplementary Fig.~\ref{fig:2}c,
where we verified that the conduction bands mainly consist of the Pt-$e_g$ states.
In other words, Ti and Pt are roughly $d^0$ and $d^6$.%
}{%
Supplementary Fig.~\ref{fig:2}b presents the first-principles band structure of pure \ce{TiSe2}.
Note that the semimetallic nature of the calculated band structure should not be taken as representative of the low-temperature phase of \ce{TiSe2}, since we did not consider the CDW order in the calculation.
The conduction band manifold consists of three isolated bands, consisting of the Ti-$t_{2g}$ orbitals.
The \ce{PtSe2}, on the other hand, has two isolated conduction bands, as shown in Supplementary Fig.~\ref{fig:2}c, consisting of the Pt-$e_{g}$ orbitals.
}


\para{Substitution vs. intercalation}
To see how Pt atoms are taken into \ce{TiSe2}, we compared the total energies of PtTi$_{71}$Se$_{144}$ for the three cases:
(1) \ce{PtSe2} and \ce{TiSe2} (i.e., \ce{PtSe2} $+$ 71 \ce{TiSe2}),
(2) Pt substituted for Ti in \ce{TiSe2} as shown in the upper panel of Supplementary Fig.~\ref{fig:2}d,
and (3) Pt intercalated into \ce{TiSe2} with a Ti defect as shown in the lower panel.
In case (3), the intercalated Pt and the Ti defect are aligned along the $c$-axis, which we found the most stable.
Calculated total energies per Pt are 0.7 and 2.5\,eV for cases (2) and (3), relative to that for case (1).
Because it was reported in the previous experimental study~\cite{chen_prb_2015} that the \ce{PtSe2} was also synthesized in the synthesis of Pt-doped \ce{TiSe2}, our observation that case (1) is the most stable is to some extent consistent with the experiment.
In addition, the occurrence of Pt-substitution is also consistent.
From now on, we shall focus on the Pt-substituted \ce{TiSe2} in the calculation.

\para{}
To see how the Pt impurity affects the electronic structure of \ce{TiSe2}, we calculated the energies of the Wannier orbitals in a Pt-doped system.
Supplementary Figs.~\ref{fig:2}e and f present the crystal structure of the Pt-substituted \ce{TiSe2} and the onsite energies of the Wannier orbitals therein, respectively.
In the Wannierization, we extracted the Pt-$e_g$ + Ti-$t_{2g}$ orbitals by disentangling their slight overlap with the valence bands in the folded band structure for the $3\times3\times2$ supercell. This was done by setting the bottom end of the outer energy window to be a bit lower than the Fermi energy (by $-0.2$\,eV).
As is clearly seen in Supplementary Fig.~\ref{fig:2}f, Pt acts as the local impurity through its different onsite energies from Ti.

\clearpage
\section{Angle-Resolved Photoemission Spectroscopy}

\begin{figure}[h]\centering
\includegraphics[width=200pt]{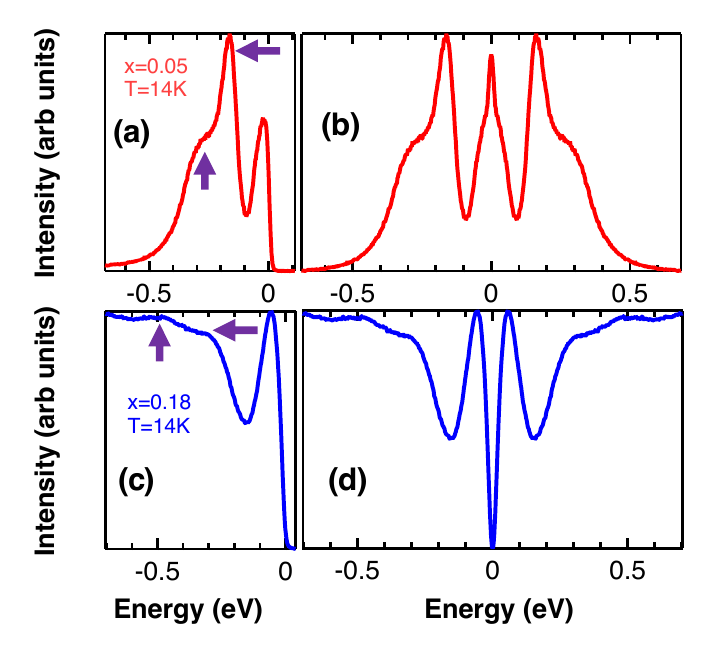}
\caption{\label{fig:ARPES-doping2}
\textbf{ARPES data for $x=0.05$ and $0.18$.}
\textbf{a}, \textbf{c}~Raw EDCs at $T \sim 14\,K$ near L point marked by the black dashed lines in Figs.~3a and 3d for $x\sim0.05$ and 0.18, respectively.
\textbf{b}, \textbf{d}~Symmetrized form of the EDCs in \textbf{a} and \textbf{c}.
}
\end{figure}

\begin{figure}[h]\centering
\includegraphics[width=140pt]{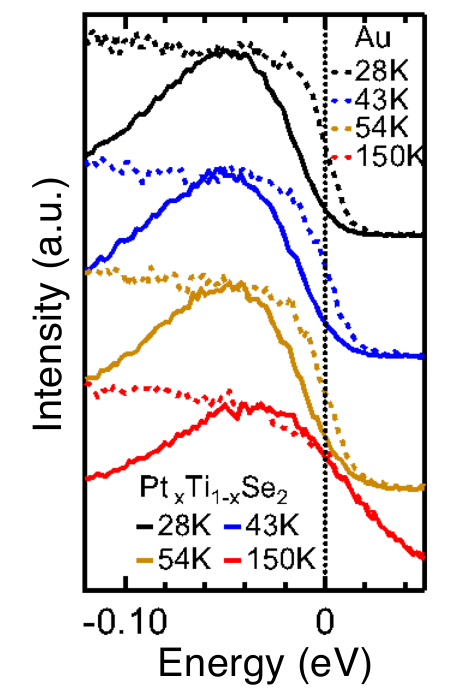}
\caption{\label{fig:ARPES-leadingedge}
\diff{}{%
\textbf{Leading-edge analysis of the ARPES data from $x\sim0.1$ sample.}
Comparison of the zoomed-in views close to $\mu$ of the leading edges of the EDCs from the $x=0.1$ sample and a polycrystalline gold sample at 28\,K, 43\,K, 54\,K, and 150\,K.
The leading edges of the EDCs of the Pt-doped sample are shifted towards negative energy with respect to that of gold at 28\,K, 43\,K, and 54\,K.
This shift and hence, the CDW energy gap is absent at 150\,K.
These are consistent with the finding that $T^* \sim 90 \, \mathrm{K}$ in Fig.~3c.}%
}
\end{figure}

\para{}
We display raw and symmetrized EDCs near L point for $x\sim0.05$ and $0.18$ in Supplementary Fig.~\ref{fig:ARPES-doping2}.
The presence of the CDW backfolded bands (indicated by purple arrows) are visible for each concentration.
However, the backfolded peaks become less pronounced for $x\sim0.18$.
Furthermore, the energy gap is absent for $x\sim0.05$ like in the case of undoped samples, while there is a finite energy gap for 0.18 similar to $x=0.1$.

\clearpage

\begin{figure}[h]\centering
\includegraphics[width=160pt]{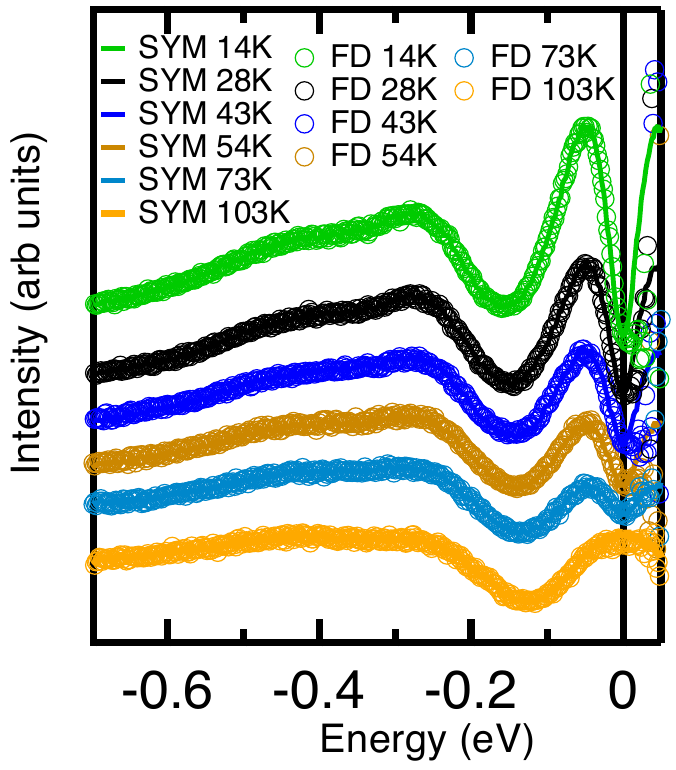}
\caption{%
\label{fig:arpes-fermi-divided}%
Comparison of temperature-dependent Fermi function-divided and
symmetrized ARPES data for $x = 0.1$ near L point.
Symmetrized data and Fermi function data are plotted on top of one another for various temperatures.
Curves are offset for visual clarity.
}
\end{figure}

\para{}
In Fig.~3b, we have used symmetrization method to approximately take out the effect of Fermi function from the EDCs.
We have also used another method for approximately taking out the effect of Fermi function from the EDCs, namely division of the EDCs by resolution-broadened Fermi function, which like symmetrization has been extensively used in the literature.
Unlike the symmetrization method, dividing EDCs by Fermi function does not invoke particle-hole symmetry approximation.
As can be found from Supplementary Fig.~\ref{fig:arpes-fermi-divided}, corresponding results are consistent with those obtained via symmetrization (Fig.~3b)—(i) the energy location of the peak of the Fermi function-divided EDC doesn't seem to change with temperature;
(ii) even at the lowest measured temperature, there is a significant spectral weight at the chemical potential;
and (iii) there is a monotonic increase in the spectral weight at the chemical potential with increasing temperature.

\clearpage
\section{Scanning Tunneling Spectroscopy}

\begin{figure}[h]\centering
\includegraphics[width=400pt]{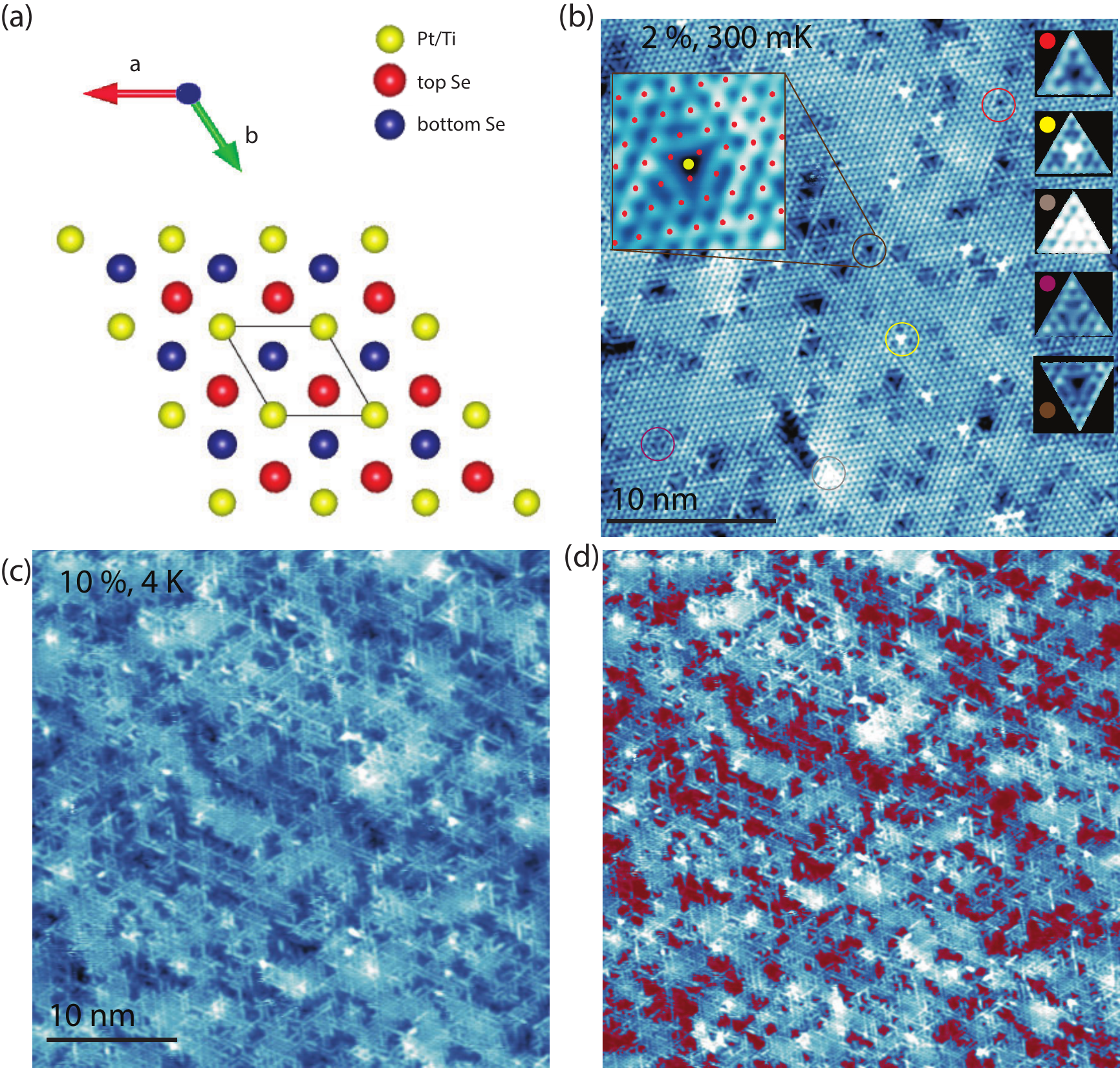}
\caption{\label{fig:SupplementaryDefect}
\textbf{a}~Sketch of the surface atomic structure.
\textbf{b}~%
30\,nm\,$\times$\,30\,nm STM topography image, recorded at 300\,mK, 400\,mV, and 250\,pA. The insets show a magnified view of the observed defects.
\textbf{c}~%
40\,nm\,$\times$\,40\,nm STM topography image, recorded at 4\,K, 500\,mV, and 150\,pA.
The black triangular shaped defects increase with the doping.
\textbf{d}~%
The same topography of \textbf{c} after masking the Pt defects.
The dark triangles occupied 34\% of the total pixels.
Since 1 pixel corresponds to 0.1\,nm and 0.35\,nm is the lattice constant, the doping is about 9.7\%.}
\end{figure}

\begin{figure}\centering
\includegraphics[width=440pt]{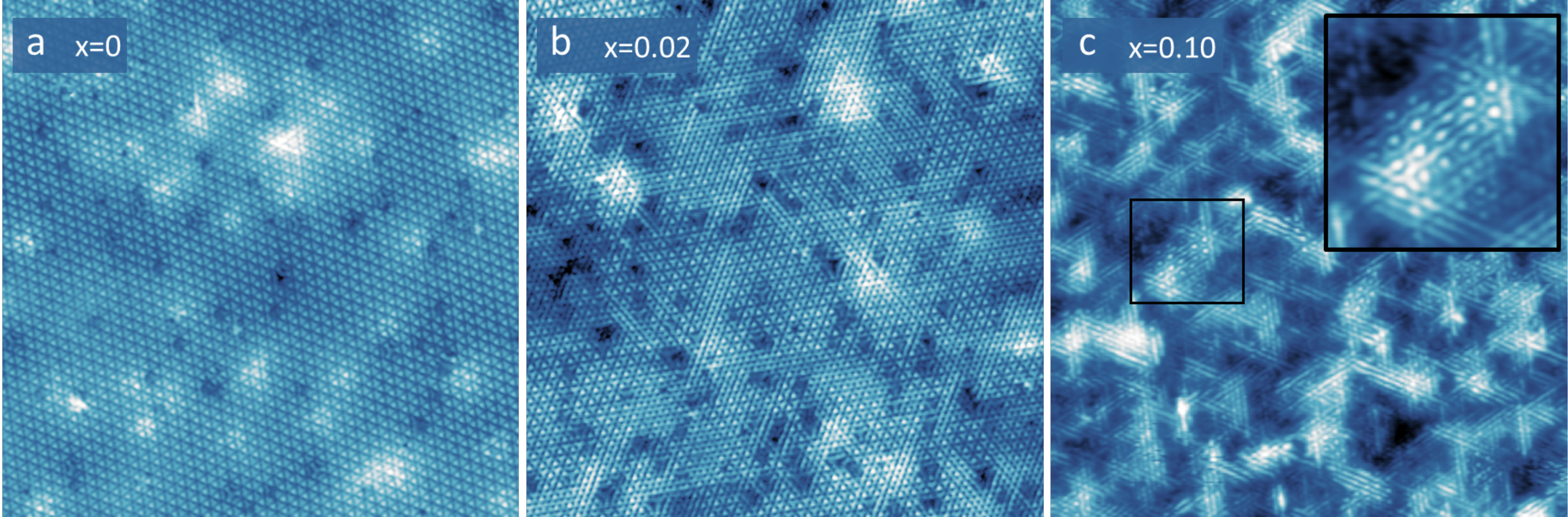}
\caption{ \label{fig:Dopingsup}
\textbf{30\,nm\,$\times$\,30\,nm STM topography images of TiSe$_2$ for different Pt doping.}
\textbf{a}~Image recorded at 4\,K, 100\,mV, and 20 pA.
\textbf{b}~Image recorded at 300\,mK, 300\,mV, and 100\,pA.
\textbf{c}~Image recorded at 4\,K, 100\,mV, and 100\,pA.
The inset is a magnified view of the are within the black box.}
\end{figure}

\begin{figure}\centering
\includegraphics[width=440pt]{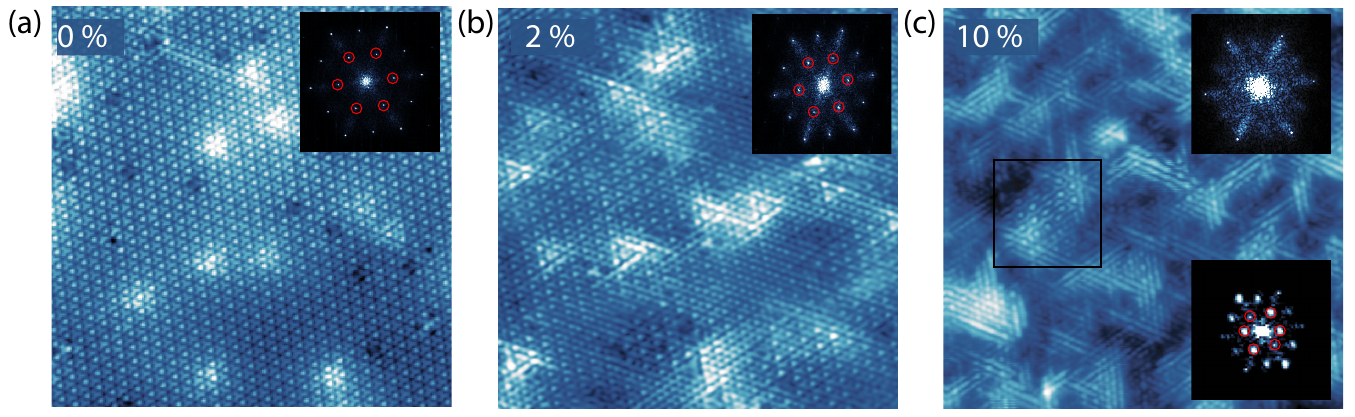}
\caption{\label{fig:FFTsup}
\textbf{20\,nm\,$\times$\,20\,nm topography images and their Fourier transforms.}
\textbf{a}~4\,K, 150\,mV, and 1\,nA, for 0\,\% Pt doping.
\textbf{b}~4\,K, 100\,mV, and 200\,pA, for 2\,\% Pt doping.
\textbf{c}~4\,K, 100\,mV, and 100\,pA, for 10\,\% Pt doping.
The top inset represents the Fourier transform of the entire area, while the bottom inset represents the Fourier transform of the area in the black box in \textbf{c}.
}
\end{figure}

\begin{figure}\centering
\includegraphics[width=500pt]{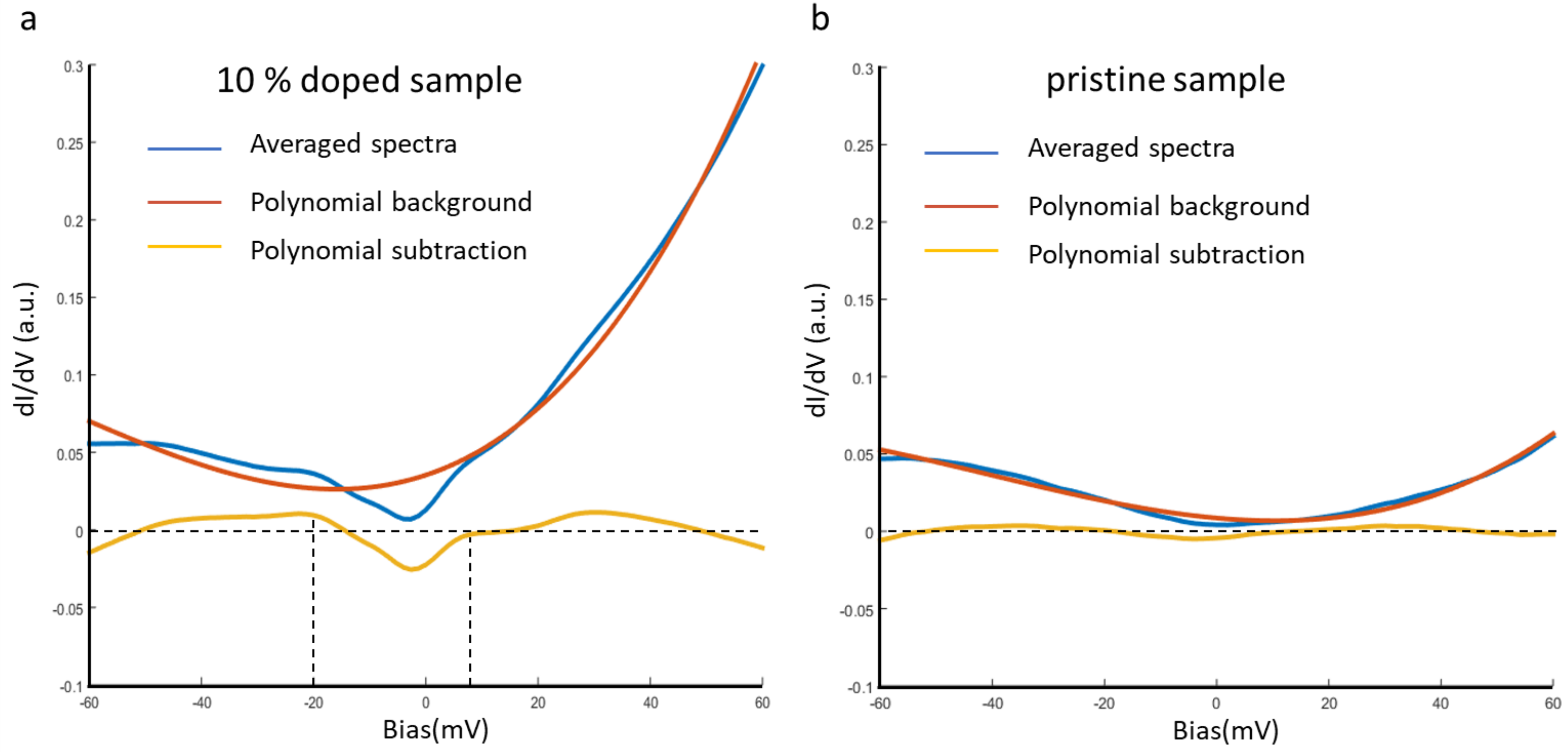}
\caption{\label{fig:polysub}
\textbf{Polynomial subtraction.}
\textbf{a}~%
Averaged spectra for the 10\,\% doped sample and third-order polynomial background.
The dashed vertical lines indicate the presence of a gap.
\textbf{b}~%
Averaged spectra for the pristine sample and third-order polynomial background.}
\end{figure}

\para{}
Using STM measurements we could identify the defects on the sample surface
(see Supplementary Fig.~\ref{fig:SupplementaryDefect}).
Particularly in the topography shown in Supplementary Fig.~\ref{fig:SupplementaryDefect}b,
for an $x = 0.02$ sample, we observed several defects.
All of them can be identified as excess Ti intercalation or Se defects, according to Refs.~\cite{Hildebrand,Novello},
except for the dark triangular shaped defects,
which increase with the doping and which are rare on the pristine sample
(see Supplementary Fig.~\ref{fig:Dopingsup}).
By tracking the atomic position, we identified these dark defects as Pt defects.
By considering the area of the image covered by Pt defects,
we could estimate a percentage of about 9.7\% doping for the sample shown in Supplementary Figs.~\ref{fig:SupplementaryDefect}c and d.

\diff{%
By increasing the Pt doping, the density of the domain wall increase as well and the CDW is more difficult to be detected, as shown by the FFT in Fig.~\ref{fig:FFTsup}.
However, the CDW is not completely suppressed, but it can still be observed in small area
(see Supplementary Fig.~\ref{fig:FFTsup}c).
}{%
}

\begin{figure}[t]\centering
\includegraphics[width=430pt]{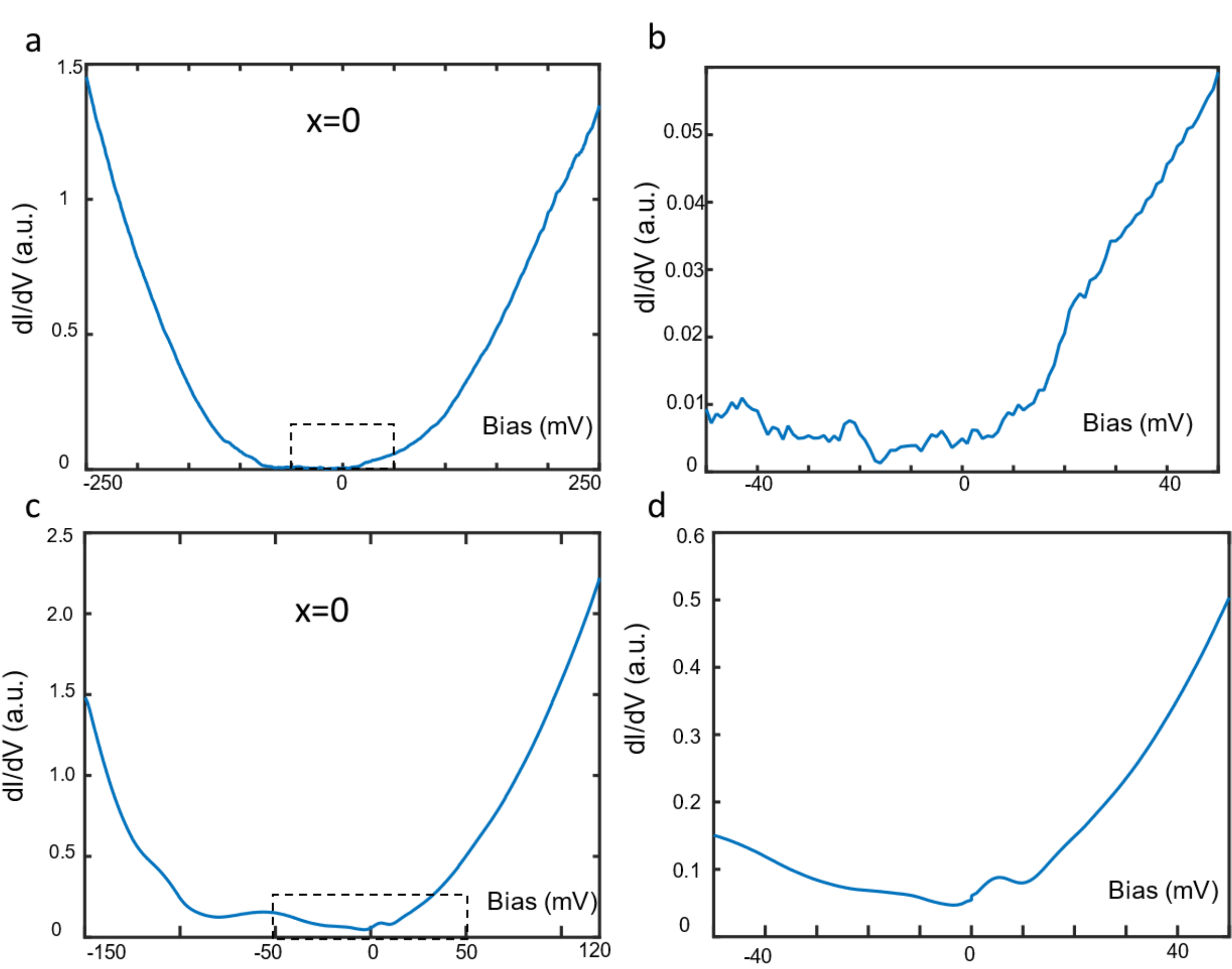}
\caption{\label{fig:Spectra}
\textbf{$\rmd I / \rmd V$ spectra for the $x=0$ sample.}
\textbf{a}~Same spectra of Fig.~4d for 0\,\% Pt doping.
\textbf{b}~Zoom-in of the dashed rectangle in \textbf{a}.
\textbf{c}~dI/dV spectra for the same sample used for \textbf{a} but on a different area.
\textbf{d}~Zoom-in of the dashed rectangle in \textbf{c}.}
\end{figure}

Supplementary Fig.~\ref{fig:Spectra} shows two spectra recorded on different areas of the same sample with Pt doping $x=0$.
Supplementary Fig.~\ref{fig:Spectra}a is the same spectra showed in Fig.~4d.
Since the shift of the chemical potential depends on the local self-doping, the same sample can exhibit small differences in the $\rmd I/ \rmd V$ spectra near Fermi energy in different areas.
However, the $\rmd I/ \rmd V$ is not flat and equal to zero.

\clearpage
\section{Inverse Fourier Transform}

The real-space image $I(\bfr)$ from the scanning tunneling microscope can be transformed to the ``momentum''-space, or $q$-space image through Fourier transform
\begin{align}
  I(\bfq) &= \int \! \mathrm{d}^{d} r \; I(\bfr) e^{i \bfq \cdot \bfr}.
\end{align}
Real-space signatures of charge density wave, which should appear as a real-space modulation in the tunneling spectra, for instance, transform into peaks in the $q$-space.
While for perfect long-range order with infinite correlation length, the peak in the $q$-space should be infinitely sharp, only bound by the resolution of the experiment.
On the other hand, when the order is imperfect---e.g. short-range order, topological defects, etc.---the peak acquires a finite width.
The real-space structure of this imperfect order parameter can be visualized through the ``selective inverse Fourier transform,'' which is a windowed Fourier transform.
\begin{align}
  I'(\bfr) \equiv \int \!\! \frac{\mathrm{d}^{d}q}{(2\pi)^d} \; I(\bfq) W(\bfq; \bfQ_1, \bfQ_2, \ldots) e^{-i \bfq \cdot \bfr}.
\end{align}
The window function $W(\bfq; \bfQ_1, \bfQ_2, \ldots)$ is a function centered around momentum $\bfQ_1, \bfQ_2, \ldots$ with a finite width (e.g. a sum of box functions $\sum_{i} \Theta(\lambda_q - |\bfq - \bfQ_i|)$, Gaussians $\sum_{i} \exp\left( -(\bfq - \bfQ)^2 / \lambda_q^2 \right)$, etc.).
See Supplementary Fig.~\ref{fig:DomainWallsSup} for an illustration of an inverse Fourier transform image. 

\begin{figure}\centering
\includegraphics[width=400pt]{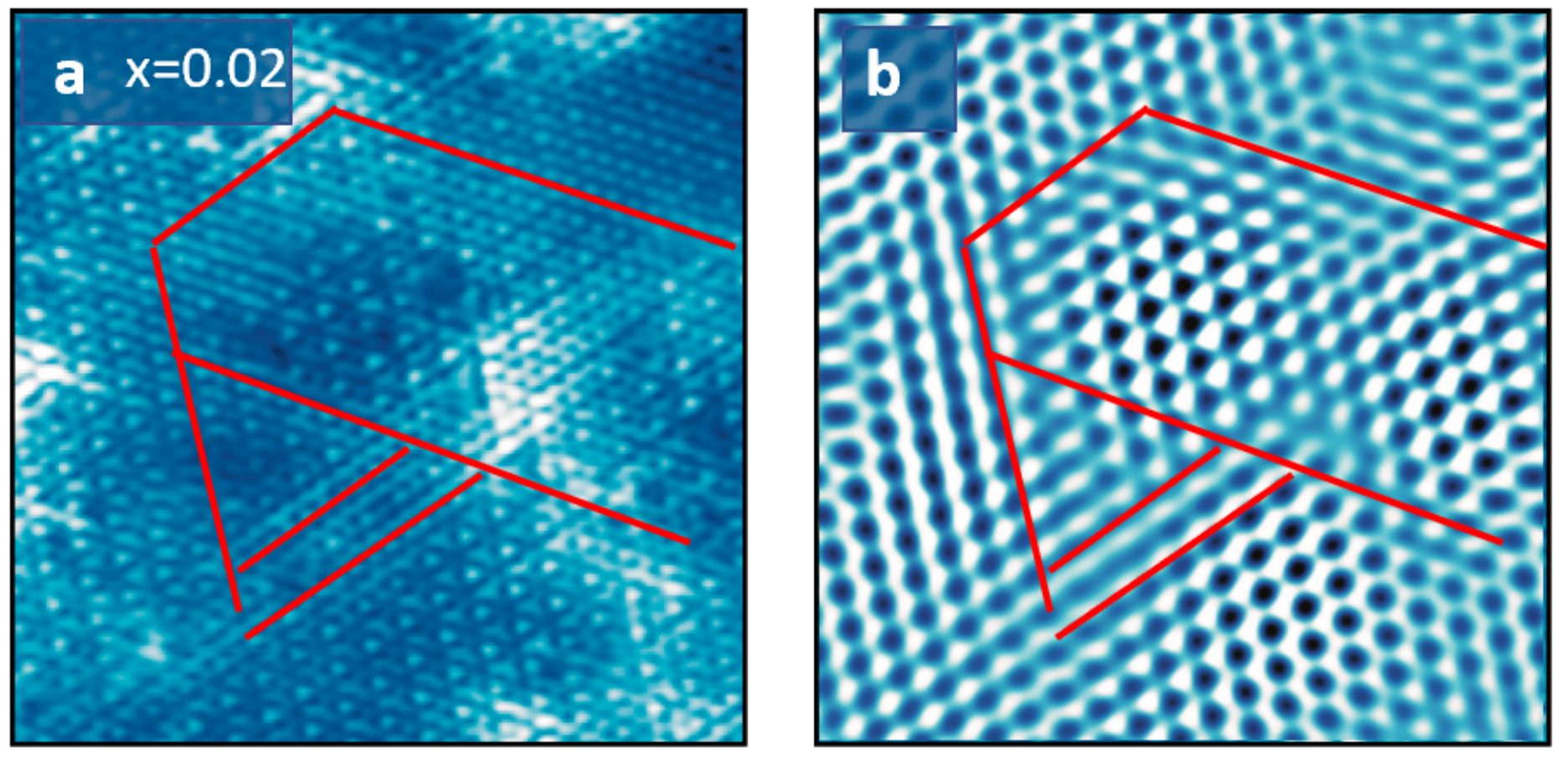}
\caption{\label{fig:DomainWallsSup}%
\textbf{CDW domain walls for the $x=0.02$ sample.}
\textbf{a}~14\,nm\,$\times$\,14\,nm topography recorded at 4\,K, $I=200$\,pA, and $V=100$\,mV. \textbf{b}~Selective inverse Fourier transform of CDW peaks calculated from \textbf{a}.
The red lines indicate the position of CDW domain walls.}
\end{figure}

\clearpage
\printbibliography[title={Supplementary References}]